\documentclass[smallestended,final]{svjour3}

\usepackage{amsmath}
\usepackage{amssymb}
\usepackage{enumerate}
\usepackage{eucal}
\usepackage{bm}
\usepackage{mathptmx}      % use Times fonts if available on your TeX system
\usepackage{natbib}
\usepackage{graphicx}
\usepackage[usenames,dvipsnames]{color}
\usepackage{hyperref}% add hypertext capabilities
\newcommand{\N}{\mathbb{N}}

\newcommand{\R}{\mathbb{R}}

\newcommand{\Prob}{\mathbf{P}}
\newcommand{\Expec}{\mathbf{E}}
\newcommand{\Varia}{\mathbf{Var}}

\newcommand{\ee}{\mathrm{e}}
\newcommand{\transp}{\mathrm{t}}

\newcommand{\bigO}{\mathit{O}}

\newcommand{\etal}{\emph{et al.}}
   
\newcommand{\ignore}[1]{}

\begin{document}

%\DeclareGraphicsExtensions{.pdf}
\DeclareGraphicsExtensions{.eps}

\title{Virus Replication as a Phenotypic Version of Polynucleotide Evolution}

\subtitle{\textrm{\small\it Dedicated to the memory of our dear colaborator and friend 
Francisco de Assis Ribas Bosco.}}

%\titlerunning{Virus Replication}  % if too long for running head

\author{Fernando Antoneli\footnotemark[1]\thanks{$\star$ corresponding author.} \and
        Francisco Bosco \and \\ 
        Diogo Castro \and Luiz Mario Janini}

\authorrunning{Antoneli, Bosco, Castro and Janini}

\institute{Fernando Antoneli \at
           Departamento de Inform\'atica em Sa\'ude,
           Universidade Federal de S\~ao Paulo, S\~ao Paulo, SP, Brazil. \\
           \email{fernando.antoneli@unifesp.br}
           \and
           Francisco Bosco \at
           Laborat\'orio de Biocomplexidade e Gen\^omica Evolutiva and 
           Departamento de Microbiologia, Imunologia e Parasitologia,
           Universidade Federal de S\~ao Paulo, S\~ao Paulo, SP, Brazil. \\
           \email{fbosco@unifesp.br}
           \and
           Diogo Castro \at
           Departamento de Medicina,
           Universidade Federal de S\~ao Paulo, S\~ao Paulo, SP, Brazil. \\
           \email{diogo.castro@unifesp.br}
           \and
           Luiz Mario Janini \at
           Departamento de Microbiologia, Imunologia e Parasitologia and
           Departamento de Medicina,
           Universidade Federal de S\~ao Paulo, S\~ao Paulo, SP, Brazil. \\
           \email{janini@unifesp.br}
}

\date{\today}
%\date{Received: date / Accepted: date} %The correct dates will be entered by the editor

\maketitle

\begin{abstract}
In this paper we revisit and adapt to viral evolution an approach based on the 
theory of branching process advanced by Demetrius, Schuster and Sigmund 
(``Polynucleotide evolution and branching processes'', \textit{Bull. Math. Biol.} 
\textbf{46} (1985) 239--262), in their study of polynucleotide evolution.
By taking into account beneficial effects we obtain a non-trivial multivariate
generalization of their single-type branching process model.
Perturbative techniques allows us to obtain analytical asymptotic expressions for
the main global parameters of the model which lead to the following rigorous results:
(i) a new criterion for ``no sure extinction'', (ii) a generalization and proof, 
for this particular class of models, of the lethal mutagenesis criterion
proposed by Bull, Sanju\'an  and Wilke (``Theory of lethal mutagenesis for viruses'', 
\textit{J. Virology} \textbf{18} (2007) 2930--2939), (iii) a new proposal for the
notion of relaxation time with a quantitative prescription for its evaluation,
(iv) the quantitative description of the evolution of the expected values in 
in four distinct ``stages'': extinction threshold, lethal mutagenesis, stationary 
``equilibrium'' and transient.
Finally, based on these quantitative results we are able to draw some qualitative
conclusions.


%\PACS{87.10.Mn \and 87.23.Kg \and 87.23.Cc \and 87.10.Ca}


\end{abstract}

%\tableofcontents

\section{Introduction}
\label{sec:INTRO}

RNA viruses exhibit a pronounced genetic diversity \citep{D1985}. 
This variability allows RNA virus to better adapt to environmental challenges as
represented by host and therapy pressures \citep{D1998}.
Due to the lack of a proofreading activity of viral RNA polymerases
(average error incorporation rate of order $10^{-4}$ per nucleotide, per
replication cycle \citep{SDH92}), short generation times and huge population numbers,
RNA viral populations may be viewed as a collection of particles bearing mutant 
genomes.

As a consequence of high mutation rates, frequencies of mutants depend not only on
their level of adaptation but on the probability of being faithfully replicated during
viral genomic RNA synthesis. 
Several studies~\citep{D1998,ES05} have looked at viral diversification processes as a
contributing cause of disease progression and of therapeutic strategies shortcomings
including vaccine trials.
It has become important to understand the process by which virus acquire diversity and
the dynamics and fluctuations of this diversity in time.
However, understanding viral evolution \emph{in vivo} has proven to be a very
unwieldy accomplishment due to the huge number of variables involved in the interplay
between viruses and their hosts.

For the last 30 years, quasi-species theory has provided a population-based framework 
to model RNA viral evolution.
Quasi-species theory is a mathematical framework that was initially formulated to
explain the evolution of macromolecules \citep{E71}. 
More recently, it has been used to describe the evolutionary dynamics of RNA viruses.
However, quasi-species theory is based on deterministic equations and so has a number
of serious drawbacks when applied to realistic experimental systems since there are
important sources of stochasticity that must be taken into account.

Traditionally, in an effort to make the viral evolution process more palpable, several
groups have addressed this subject from different points of view.
There is a substantial amount of publications that studied virus populations during
their evolution in experimental settings, for instance, cell cultures \citep{ELM06},
by challenging the virus with population bottlenecks \citep{OBD10,OBANED10}, or the 
introduction of antiviral drugs \citep{CCA01}, including mutagens, or another competing
viral population.
Several other groups have studied the process of viral evolution away from the bench,
using computational and mathematical tools~\citep{WWOLA01,BSW07}.
The main lesson learned from these efforts is that in order to escape from
oversimplifying the interplay between virus and hosts, one needs to take into account
a few hard rules based on the experimental data that has been generated by the 
community of investigators addressing viral evolution. 

In this paper we revisit and adapt to viral evolution an approach based on the
theory of branching process advanced by~\cite{DSS85} in their study of polynucleotide
evolution.
In fact, almost any stochastic model of asexual replication has an underlying branching
process, which remains implicit and undefined in most of the studies in this subject.
For instance, \citep{MLPED03,AM08,ALM09,CACM11,CCMA11,C11} have investigated 
probabilistic models in the form of a linear mean field approximation without explicitly
defining the underlying stochastic process modeling the microscopic dynamics of particle
replication.
However, there are a few of them~\citep{DSS85,HRWB02,W03} that take a different path
and explicitly define the stochastic process in order to bring the mathematical theory
of branching processes to bear.
This attitude has some virtues, since it provides powerful tools, that have been
perfected in the past several decades, allowing one to extract quantitative results
on a rigorous basis in clear conceptual framework.

In the present work we pursue the same path as~\citep{DSS85,HRWB02,W03}, by putting
more emphasis on the study of an explicit generating function of the underlying
branching process and applying the mathematical tools as a means to get new insights 
and perspectives on the dynamics of evolving virus populations. 
Before setting up our mathematical framework let us give a broad overview of the scenery
that we concieve as the context of the models described here.

Viral infections may result in different clinical outcomes which are consequences of
interactions between the virus and its host. 
The constant arms race between hosts and virus can either eliminate or perpetuate viral
infections. 
Some viral diseases are self limited ending with the eradication of the infecting virus
while others lead towards persistent infections.

During its adaptation to new hosts virus populations will show an intense fluctuation in
size and structure in response to the selective pressures imposed by the host defenses.
Soon after the infection virus particles start to colonize the new environment and will
seek within the host for the cell types that will better support the production of an
exuberant progeny. 
At this moment and with the total absence of an immune response the viral population may
experience an \emph{exponential growth}.
This growth is clinically reflected by the appearance of acute symptoms and the emergence
of a \emph{peak of viremia} detected in the peripheral blood.  
The emergence of selective pressures represented by the combination of components present
in the innate and adaptive responses of the immune system will cause a drop in the viral
load to levels well below the initial viremia value.
At this point the virus and host defenses may reach a \emph{stationary ``equilibrium''} 
referred to as the \emph{viral set point} in some chronic infections as the one caused by
the human immunodeficiency virus (HIV).
On the other hand, if the burden imposed by the host pressures overcomes the virus
survival capacity, the viral population will become \emph{extinct}.
The initial exponential growth and subsequent population size reduction experienced by 
the incoming virus can be referred to as the \emph{transient phase} of a viral growth
curve. 
The transient phase will represent the virus population fluctuations during the period
from its introduction in the new host until the moment the viral set point is reached.
In persistent infections without major oscillations, in the host selective pressures or 
in virus adaptability, virus populations will show an \emph{asymptotic behavior} and will
perpetuate in a stationary equilibrium regime.
In virus with genome molecules represented by RNA which are able to tolerate high
mutational rates, the stationary equilibrium is actually a balance between the two major
forces in virus evolution: mutation and selection.
As a matter of fact, the transient phase of a virus growth has been also called the
\emph{recovery time} as an allusion to virus populations recovering from bottleneck
passages that correspond to drastic population size reductions.
When selective pressures against the incoming virus are too strong for the virus to 
couple with them the stationary phase of the virus growth is never attained and the
asymptotic behavior is absent.
However, changes that will cause a recrudescence in the host selective pressures may
dislocate virus from the stationary equilibrium towards its extinction.
Therapeutic strategies as antiviral drugs and therapeutic vaccines may precipitate virus
populations in regions of instability represented by hostile adaptations marked by
subsequent drops in virus replication capacity and progeny sizes.
This phase of viral growth named here as \emph{extinction threshold} represents a moment
of uncertainty when the virus extinction is strongly possible but unpredictable.
Moving forward, when virus populations are unable to counteract the deleterious effects 
of all natural and chemical pressures combined, and average virus progeny drops to a 
value bellow one viable particle, the process called the \emph{lethal mutagenesis} is
started.
At this moment the virus will become extinct almost certainly.
Virus extinctions can be seen during self limited infections even before an asymptotic
behavior is reached or when chronic infections are cleared due to optimization of host
pressures or to the addition of antiviral chemicals.

Once the mathematical framework is established, the scenary described above raises
several questions about the evolution of a viral population within its host can be posed
and answered in a precise way.
In this work we shall address the following questions:
\begin{enumerate}[(i)]
\item What is the average number of particles in the population in each generation?
\item What is the probability of extinction of the population within its host?
\item How does the ``transition'' to extinction happens?
\item When the population does not become extinct, what is the asymptotic behaviour?
\item How long it takes for the population to reach its asymptotic state?
\end{enumerate}
In order to answer these questions we propose a non-trivial multivariate generalization
of the single-type branching process of~\cite{DSS85}.

By employing perturbative techniques we are able to obtain analytical expressions for
the main global parameter of the model, namely the average growth rate, represented 
here by the \emph{malthusian parameter} of the branching process.
Our results corroborate with the study of~\cite{BSW07} by showing that the sufficient
condition for lethal mutagenesis involves mutational and ecological aspects.
They arrived at a conjectural criteria for lethal mutagenesis by
a heuristic and intuitive approach of great general applicability.
%{\color{Red}
For the particular class of models considered here it is possible to state and 
rigorously prove an \emph{extinction criterion} which is, under a proper interpertation,
equivalente to the \emph{lethal mutagenesis criterion} of~\cite{BSW07}.
In other words the branching process formalism can provide a new and interesting
perspective on this problem.
%}
We obtain a new \emph{criterion for non-extinction} of a viral population and
describe four distinct ``stages'' of evolution of a RNA virus populations: extinction
threshold, lethal mutagenesis, stationary ``equilibrium'' and decay of the temporal
auto-correlation.

\section{Phenotypic Models for Viral Evolution}
\label{sec:PMVE}

In this section we describe a model for viral evolution that is naturally represented
by a multivariate branching stochastic process generalizing, in a non-trivial way, the 
single-type branching process studied by~\cite{DSS85}.
For the sake of motivation we start by recalling a probabilistic model introduced 
by~\cite{LEDM02}.

We interpret the notion of mutation probability as a general effect of probabilistic
nature with direct impact on the replication capability of individual viral particles, 
considered here as a measure of the particle's fitness characterizing its phenotype.
This effect is summarized by the definition of a stationary probability 
distribution which is used to set up a Galton-Watson branching process~\citep{WG1874}
for the temporal evolution of the viral population. 
This probability distribution gives appropriate parameters to classify the asymptotic
behavior of the viral population and to describe some of the non-equilibrium properties
of the model. 

In other publications on the same subject the concept of mutation is extensively used as
the cause of replication capacity change.
Understanding that those changes constitute an observable output due to many different
factors (of genetic and non-genetic nature), we prefer to use the general term
``effect'' over the replication capacity to characterize the three possible changes 
(deleterious, beneficial and neutral) that may happen with the viral particle when it 
replicates.

\subsection{The Probabilistic Model}

A number of viral infections starts with the transmission of a relatively small number
of viral particles from one organism to another one. 
The initial viral population starts replicating constrained by the unavoidable
interaction with the host organism and evolves in time towards an eventual equilibrium.
Each particle composing the population replicates in the cellular context that may
differ from cell to cell.
Moreover each particle has different replication capabilities due to the natural genomic
diversity found in viral populations in general.
Therefore, it is reasonable to consider the viral population as a set of particles
divided in groups of different replication capabilities measured in terms of the number
of particles that one particle can produce.
The models considered here do not take into account any information about the genomic
diversity of any replicating class and therefore it should be classified as phenotypic
model.  

Consider that the whole set of particles composing the viral population replicates at
the same time in such a way that the evolution of the population is described as a
succession of discrete viral generations. 
This assumption crucially depends on the clear definition of the time needed for a
particle to replicate, referred by virologists as \emph{generation time}. 
As it depends on the cell environment it is clear that this time period may vary from
particle to particle, replicating in different cells, in such a way that a meaningful
concept is a distribution of replication times with a possible well defined mean value.
The dispersion of the replication times can be considered small if we restrict ourselves
to homogeneous cell populations.
Under these conditions, one may consider that no particle can be part of two successive
generations, that is, the generations are \emph{discrete} and \emph{non-overlapping}.

Suppose that a population of viruses that start evolving from an initial
set of particles, which is partitioned into \emph{classes} according to their
\emph{mean replication capacity}, that is, for each integer $r=0,\ldots,R$ 
there is a class of particles labeled by $r$ and a random variable assuming non-negative
integer values whose probability distribution $t_r(k)$ satisfies $\Expec(t_r)=r$ and
$t_0(k)=\delta_{k0}$.

\ignore{
Typical examples of \emph{replication capacity distributions} are: 
(i) the delta distribution $t_r(k)=\delta_{kr}$,
(ii) the Poisson distribution $t_r(k)=\tfrac{r^k \ee^{-r}}{k!}$ if $r \geqslant 1$
and $t_0(k)=\delta_{k0}$,
(iii) the geometric distribution $t_r(k)=\tfrac{1}{r}(1-\tfrac{1}{r})^{k}$
if $r \geqslant 2$ and $t_r(k)=\delta_{kr}$ for $r=0,1$.
Note that in the first case the replication capacity is sharply concentrated on
the mean value -- it is deterministic.
}

Inasmuch as the process of replication is controlled by chemical reactions involving
specific enzymes and the template, it is reasonable to assume a mean bounded 
replication capacity per particle that is possibly typical for each specific virus.
Hence there is a \emph{maximum mean replication capacity} $R$ imposed by the
natural limiting conditions under which any particle of the population replicates.

In the process of replication of a viral particle errors may occur at each replication 
cycle in the form of point mutations with possible impact on the replication capacity
of the progeny particles.
Due to the intrinsic stochastic component of chemical reactions it is natural to treat
this point mutational cause as probabilistic.
Another possible cause of change in the replication capability in the viral offspring
is clearly related to the cellular environment where the replication process takes 
place. 
As a result the time evolution of viral populations should be viewed as a physical
process strongly influenced by stochasticity.
Therefore, the combined action of genetic and non-genetic causes may produce basically
three types of replicative effects with associated probabilities, at the particles
scale, applicable to every single replication event:
\begin{itemize}
\item \emph{deleterious effect $d$}: the mean replication capacity of the copied 
      particle decreases by one. Note that when the particle has capacity of
      replication equal to $0$ it will not produce any copy of itself on the average.
\item \emph{beneficial effect $b$}: the replication capacity of the copy increases by
      one. If the mean replication capacity is already the maximum allowed then the
      mean replication capacity of the copies will stay the same. 
\item \emph{neutral effect $c$}: the mean replication capacity of the copies remain
      the same as the mean replication capacity of the parental particle.
\end{itemize}
The only constraint these numbers should satisfy is $b+c+d=1$.
In the case of \emph{in vitro} experiments with homogeneous cell populations the
parameters $c$, $d$ and $b$ may be considered essentially as mutation probabilities.

The assumption that $b$ is very small when compared with $d$ and $c$ is justified by
several experimental results.
The frequencies between beneficial, deleterious and neutral mutations appearing in a
replicating population have been already measured by prior 
studies \citep{MGME99,IS01,KL03,O03,SME04,CIE07,EWK07,PFCM07,RBJFBW08}.
Taking their results together, it is reasonable to conclude that beneficial mutations
could be as low as $1000$ less frequent than either neutral or deleterious mutations. 
As a result, the viral population would be submitted to a large number of successive
deleterious and neutral changes and a comparatively small number of beneficial changes.
The particular case where there are no beneficial effects in time ($b=0$) is interesting
not only because of biological reasons but also due to a mathematical property, namely,
the spectral problem associated to its mean matrix is ``completely solvable''.
This property will be crucial for us to implement the perturbative expansions in the
general case where $b \neq 0$.

\subsection{The Multitype Branching Process}

Instead of working directly with a probabilistic model we will introduce a new 
generating function, which is a non-trivial multivariate generalization of the
single-type branching process proposed by~\cite{DSS85}.
For full presentations of the theory of branching processes see 
\citep{H63,AN72,K02}.
Brief accounts of the theory branching process, highlighting the main results 
we need here, are provided by~\cite[Sec. 2]{DSS85} and~\cite{ABCJ11}.

In order to find our probability generating function we observe that the replication
process is simply a \emph{Bernoulli trial} with three possible outcomes: a newly
produced particle may have endured a deleterious, neutral or beneficial effect.
Therefore, the offspring distribution should be a \emph{trinomial distribution} 
(see~\cite{F68}) and the probability generating function of the phenotypic
model is $\bm{f}=(f_0,f_1,\ldots,f_R)$ with components
\begin{equation} \label{EQ:genfunc2}
 f_s(z_0,z_1,\ldots,z_R) = \sum_{k=0}^\infty \,t_s(k)\, (dz_{s-1}+cz_s+bz_{s+1})^k \,.
\end{equation}
with ``consistency conditions'' $f_0(z_0,z_1,\ldots,z_R)=1$ and $z_{R+1}=z_{R}$.

\ignore{
\begin{equation} \label{EQ:genfunc2}
\begin{split}
 f_0(z_0,z_1,\ldots,z_R) & = 1 \\
 f_1(z_0,z_1,\ldots,z_R) & = \sum_{k=0}^\infty \,t_1(k)\, (dz_0+cz_1+bz_2)^k \\
 f_2(z_0,z_1,\ldots,z_R) & = \sum_{k=0}^\infty \,t_2(k)\, (dz_1+cz_2+bz_3)^k \\
                        & \vdots \\
% f_{R-1}(z_0,z_1,\ldots,z_R) & = \sum_{k=0}^\infty 
%\,t_{R-1}(k)\, (dz_{R-2}+cz_{R-1}+bz_R)^{k} \\
 f_R(z_0,z_1,\ldots,z_R) & = \sum_{k=0}^\infty \,t_R(k)\, (dz_{R-1}+(c+b)z_R)^k
\end{split}
\end{equation}
}

Note it is trivial to further generalize the model in order to include 
``higher order replicative effects'' that changes the mean replication capacity 
by $\pm 2, \pm 3, \ldots$ and thus replacing the trinomial distribution by a 
multinomial distribution.
It is also worth to mention that there are other possible variations of these models
that share the same essential properties and are more adequate in different contexts:
\begin{description}
\item[\bf With Zero Class:] In this version the model naturaly fits the symmetry of 
      the binomial distribution since it contains the particles of class $r=0$ which
      are generated by the particles from class $r=1$.
\item[\bf Without Zero Class:] In this variation, the particle class $0$ is omitted 
      and thus the probability generating function has $R$ variables and $R$
      components: set the variable $z_0=1$ and omit the first component $f_0$.
      Particles of class $r=1$ undergoing a deleterious change are eliminated in
      the next generation.
      In this formulation the model is \emph{positively regular}.
\end{description}

With the convention described above, it is easy to see that in the one-dimensional case
with $b=0$ one obtains the following generating function
\begin{equation} \label{eq:DSS}
  f(z)~=~\sum_{k=0}^\infty \,t(k)\, ((1-c)+cz)^k~=~\sum_{k=0}^\infty \,t(k)\, (1-c(1-z))^k \,.
\end{equation}
This is exactly the generating function of the single-type model proposed 
by~\cite[p. 255, eq. (49)]{DSS85} for the evolution of polynucleotides.
In their formulation, $c=p^\nu$ is the probability that a given copy of a polynucleotide
is exact, where the polymer has chain length of $\nu$ nucleotides and there is a fixed
probability $p$ of copying a single nucleotide correctly. 
The replication distribution $t$ provides the number of copies a polynucleotide yields
before it is degraded by hydrolyses.

\subsection{Evolution of the Mean Values}

We shall introduce the notation $Z_0^r=1$ for the condition $\bm{Z}_0=\bm{e}_r$,
which is the initial population consisting of one particle of class $r$ and no
particles of other classes.
Thus $\Prob(\bm{Z}_1=\bm{i}|Z^r_0=1)=\zeta_r(\bm{i})$.
A basic assumption in the theory of branching processes is that all the first
moments are finite and that they are not all zero. 
Then one may consider the \emph{mean evolution matrix} or the
\emph{matrix of first moments} $\bm{M}=\{M_{ij}\}$ which describes how the averages
$\langle \bm{Z}_{n} \rangle$ of the sub-populations of particles in each replication 
class evolves in time:
\begin{equation} \label{eq:MEANEVOL1}
 \langle \bm{Z}_{n} \rangle=\bm{M}^n\,\bm{Z}_0 
 \quad\text{or}\quad \langle \bm{Z}_{n} \rangle=M\langle \bm{Z}_{n-1} \rangle \,.
\end{equation}
The \emph{mean matrix} $\bm{M}=\{M_{ij}\}$ is defined as
\[
 M_{ij}=\Expec(Z_1^i|Z_0^j=1) \qquad\text{or}\qquad
 M_{ij}=\dfrac{\partial f_j}{\partial z_i}(\bm{1}) \,,
\]
for $i,j=0,\ldots,R$ and $\bm{1}=(1,1,\ldots,1)$.

\ignore{
Denoting by $\bm{f}'$ the jacobian matrix of $\bm{f}$ one may write
\[
 \bm{M}=\bm{f}'(\bm{1})\,.
\]
}

In order to compute the mean matrix of \eqref{EQ:genfunc2} we observe that
only the partial derivative of a component $f_r$ with respect to $z_{r-1}$,
$z_r$ and $z_{r+1}$ are non-trivial, the others are zero.
Therefore, the mean matrix is
\begin{equation} \label{eq:MEANGENERAL}
\bm{M}=\begin{pmatrix}
 0 & d &  0 &  0 & \cdots & 0 & 0 \\
 0 & c & 2d &  0 & \cdots & 0 & 0 \\
 0 & b & 2c & 3d & \cdots & 0 & 0 \\
 0 & 0 & 2b & 3c & \cdots & 0 & 0 \\[-2mm]
\vdots & \vdots  & \vdots & \vdots & \ddots & (R-1)c & Rd \\
 0 & 0 &  0 & 0  & \cdots & (R-1)b & R(c+b)
\end{pmatrix}
\end{equation}
Thus showing that our generating function indeed provides a non-trivial 
(i.e., non-trianglar) generalization of \eqref{eq:DSS}.
Interestingly, the mean matrix $\bm{M}$ provided by this class of models is a
\emph{tridiagonal matrix}, which is an ubiquitous type of matrix appearing in 
several fields ranging from statistical signal processing \cite{G06}, information 
theory \cite{GS58}, lattice dynamical systems \citep{ADGW05}.

The mean matrix obtained above coincide with some of the matrices in the linear 
mean field approximation studied in \cite{MLPED03,AM08}, \linebreak
\cite{ALM09,CACM11,CCMA11,C11}.
However, it is important to stress that these works consider that the
\emph{total population} $\bm{Z}_n$ evolves by multiplication by the mean matrix,
while here it is the evolution of the \emph{average value} $\langle \bm{Z}_{n} \rangle$
that is described by the mean matrix.
However, the mean matrix $\bm{M}$ depends on the probability distributions 
$t_r(k)$ only through its mean value $r$.
Therefore, there are infinitely many distinct branching processes with the
same mean matrix and the only way to tell them apart is by looking at their
fluctuations or second moment properties.

Let $\varrho(\bm{M})$ denote the \emph{spectral radius} of the mean matrix $\bm{M}$,
that is, if $\lambda_1 \leqslant\cdots\leqslant\lambda_R$ are the eigenvalues of
$\bm{M}$ then $\varrho(\bm{M})=\lambda_R$.
Since $\bm{M}$ is a non-negative matrix, it has at least one largest non-negative
eigenvalue which coincides with its spectral radius (see~\cite{G05} or~\cite{M00}).
When the largest eigenvalue is positive we shall call it, following~\cite{K02}, 
the \emph{malthusian parameter} $m=\varrho(M)=\lambda_R$ of the branching process 
(see also~\cite{JKS07}).

The main classification result of multitype branching processes, in the 
\emph{positively regular} case, states that there are only three possible 
regimes (see~\cite{H63} or~\cite{AN72}): \emph{sub-critical} ($m < 1$),
\emph{super-critical} ($m > 1$) and \emph{critical} ($m=1$).
Note however, that when $b=0$, the corresponding branching process is \emph{not}
positively regular.
Nevertheless, there is a generalization of the classification of multitype branching
processes, due to Sevastyanov (see~\cite{H63,J70}), that allows us to include the case
$b=0$ within the same three regimes as above.

\section{Malthusian Parameter of the Phenotypic Model}
\label{sec:MPPM}

In this section we shall employ the perturbation theory for eigenvalues and
eigenvectors in order to study the average dynamics of the phenotypic model,
with $b$ small.
In order to maintain the main flow of the text we have omitted most of the computations
and estimates which have been placed in the appendix for the convenience of the reader.
See also~\cite{W65} or~\cite{TM74} for more details.

Methods of spectral perturbation have been used in the study of evolution of
macromolecules \citep{SS82,TM74}, where the the starting point is a diagonal matrix
and the perturbation adds other off-diagonal elements. 
Our approach is similar, but our starting point is an upper triangular matrix and
the final matrix is tridiagonal.

In order to implement this program it is necessary to know all the eigenvalues of
the unperturbed matrix and their associated left and right eigenvectors.
The idea is to write the mean matrix \eqref{eq:MEANGENERAL} as sum of a matrix
whose spectral problem is exactly solvable with a ``perturbation matrix'' depending on
a small parameter, in such a way that the perturbation vanishes as the parameter goes
to zero.

Let $\bm{M}$ be the mean matrix of~\eqref{eq:MEANGENERAL} the phenotypic model.
Then, since $c=1-b-d$ we may write
\[
 \bm{M}=\bm{M}_0+b\,\bm{P} \,,
\]
where $\bm{M}_0$ is the mean matrix of the simple phenotypi model:

\begin{equation} \label{eq:MEANBASIC}
\bm{M}_0=\begin{pmatrix}
 0 & d     &  0     & \ldots & 0 \\
 0 & (1-d) & 2d     & \ldots & 0 \\
 0 & 0     & 2(1-d) & \ldots & 0 \\
 0 & 0     &  0     & \ldots & 0 \\
\vdots     & \vdots & \vdots & \ddots & Rd \\
 0 & 0 & 0 & \ldots & R(1-d)
\end{pmatrix}
\end{equation}
and $P$ is the matrix
\begin{equation} \label{eq:PERTURBATION}
\bm{P}=\begin{pmatrix}
 0 &  0 &  0 &  0 &  0 & \ldots &  0 & 0 \\
 0 & -1 &  0 &  0 &  0 & \ldots &  0 & 0 \\
 0 &  1 & -2 &  0 &  0 & \ldots &  0 & 0 \\
 0 &  0 &  2 & -3 &  0 & \ldots &  0 & 0 \\
 0 &  0 &  0 &  3 & -4 & \ldots &  0 & 0 \\
 \vdots &  \vdots &  \vdots & \vdots & \vdots & \ddots & \vdots & \vdots \\ 
 0 &  0 &  0 &  0 & 0  & \ldots & -(R-1) & 0 \\
 0 &  0 &  0 &  0 & 0  & \ldots &  (R-1) & 0
\end{pmatrix} \,.
\end{equation}
It is easy to see that $\bm{M}$ is the mean matrix of the phenotypic model when
one sets $b=0$ and $c=1-d$ in the generating function \eqref{EQ:genfunc2}.
An interesting feature of this particular case is that its spectral problem is
completely solvable as functions of the parameters $d$ and $R$.
The eigenvalues $\lambda_r^0$ of the mean matrix $\bm{M}_0$ are:
\[
 \lambda_r^0=rc=r(1-d) \qquad r=0,\ldots,R \,.
\]
In particular, the \emph{malthusian parameter} is the largest positive eigenvalue
\begin{equation} \label{eq:MALTHPAR}
 m^0=\lambda_R^0=Rc=R(1-d) \,.
\end{equation}
Also, note that the operator norm of $\bm{P}$ is $\|\bm{P}\|=2(R-1)$.

Now we write the eigenvalue $\lambda_r$ of $\bm{M}$ as a function of the parameter $b$,
expanded as a power series
\[
 \lambda_r = \lambda_r^0 + \lambda_r^1 b + \lambda_r^2 b^2 + \ldots \,,
\]
where $\lambda_r^0$ is the corresponding eigenvalue of $\bm{M}_0$.
Clearly, $\lambda_r(b)\to\lambda_r^0$ as $b\to 0$.
The higher order perturbation terms $\lambda_r^i$, ($i=1,2,3,\ldots$) 
are written in terms of all the eigenvalues $\lambda_s^0$ ($s=0,\ldots,R$) 
of $\bm{M}_0$ and their associated left and right normalized eigenvectors 
$\bm{v}_s^0$ and $\bm{u}_s^0$.

We are interested in the malthusian parameter $m$ of the matrix $\bm{M}$:
\[
 m = m^0 + m^1 b + m^2 b^2 + \ldots \,,
\]
where $m^0=(1-d)R$.
A lengthy calculation gives the second order expansion of the malthusian parameter $m$:
\begin{equation} \label{eq:PERTEXPAN}
\begin{split}
 m = & (1-d)R + \dfrac{(R-1)\,d}{1-d}\,R\,b \\
     & -\dfrac{(1+d\,R/2)(R-1)^2 d}{(1-d)^3} \,R\,b^2 + \bigO(b^3) \,.
\end{split}
\end{equation}

\begin{figure}[!hbt] 
\begin{center}
 \includegraphics[scale=0.25,angle=0]{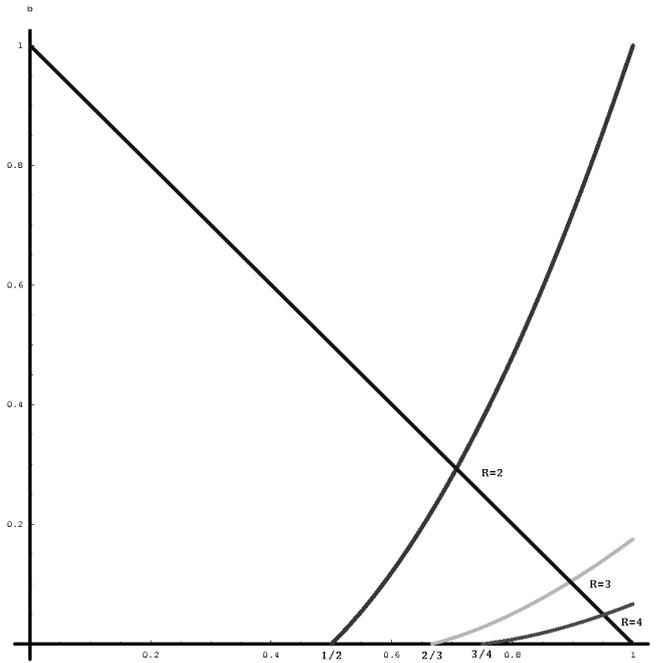}
 \caption{\label{fig:GENERAL1} Parameter space $(d,b)$ of the phenotypic model 
          with the critical curves $m=1$ for $R=2,3,4$.} \vspace{-5mm}
\end{center}
\end{figure}

The parameter space of the model is $\{(b,d,R)\in\R^2\times\N~|~b+d\leqslant 1\}$.
If we forget the discrete parameter $R$, then this set can be
identified with the triangle $\{(0,0),(0,1),(1,0)\}$ in the $(d,b)$-plane
(see FIG.~\ref{fig:GENERAL1}).

Now we can make some quantitative predictions about the behaviour of the model.

One can localize in the parameter space $(b,d)$ each of the three regimes of the
phenotypic model (\emph{sub-critical}, \emph{super-critical}, \emph{critical})
by considering, for each fixed $R$, the \emph{critical curve} $m(b,d,R)=1$ 
(see FIG.~\ref{fig:GENERAL1}).
The \emph{region of sub-criticality} (for each fixed $R$) lies on the right side of 
the critical curve and the \emph{region of super-criticality} (for each fixed $R$) 
lies on the left side of the critical curve.

\ignore{
In the particular case where $b=0$ we have the following.

\begin{theorem} \label{thm:BASIC}
The simple phenotypic model has three distinct regimes.
\begin{enumerate}[(i)]
\item If $m=R(1-d)<1$ then the branching process is \emph{sub-critical}.
      That is, with probability $1$, the virus population becomes extinct
      in finite time.
\item If $m=R(1-d)>1$ then the branching process is \emph{super-critical}.
      That is, with positive probability, the virus population survives
      and grows indefinitely at an exponential rate proportional to $m^n$
      when $n\to\infty$.
\item If $m=R(1-d)=1$ then the branching process is \emph{critical}.
      That is, with probability $1$, the virus population becomes
      extinct but this may take an infinite time to happen. 
\end{enumerate}
\end{theorem}
\begin{proof}
Since the simple phenotypic model does not have a singular component the main result
about the classification of a multitype branching process without singular components
(see~\cite{H63} and~\cite{J70}) and equation \eqref{eq:MALTHPAR} give the result. \qed  
\end{proof}
}

Another interesting feature of the parameter space is related to
the points of intersection of the critical curve $m(b,d,R)=1$ with 
the boundary curves $b+d=1$ and $b=0$.
The points of intersection with the boundary curve $b=0$ are give by
$d_c(R)=1-1/R$ and the points of intersection with the boundary curve
$b+d=1$ have non-zero coordinates and can be written as $(1-b_c,b_c)$
where $b_c(R)$ is the $b$-coordinate and $1-b_c(R)$ is the $d$-coordinate.
For instance, it is easy to find the exact value of $b_c$ for $R=2$:
$b_c(2)=1-\tfrac{1}{\sqrt{2}}\cong 0.2928$.
This calculation gives an important prediction: if $b>0.2929$ then the population
can not become extinct by increasing $d$ towards its maximum value ($\cong 0.7070$), 
with probability going to $1$ as $R$ becomes large.

More generally, one has the following \emph{criterion of no sure extinction}.
Consider the phenotypic model with parameters $(b,d,R)$, with $R$ large enough.
If $b\neq 0$ is sufficiently small and (up to order $\bigO(b^2)$)
\begin{equation}  \label{thm:NOEXTINCTAPPROX}
 b \, R(R-1)^2 \geqslant 1
\end{equation}
then, asymptotically almost surely, the population can not become extinct by 
increasing the deleterious probability towards its maximum value.
In other words, when the inequality \eqref{thm:NOEXTINCTAPPROX} is satisfied
the process can have only one regime, namely, the supercritical, which
renders the model completely trivial, since there is no transition.

This criterion follows from the approximation of $b_c(R)$ that can be
obtained from the truncated perturbation expansion of the malthusian parameter.
Solving the equation $m(b,1-b,R)=1$ for $b$ in terms of $R$ gives
\begin{equation} \label{eq:BAPPROX}
 b_c(R) \approx \dfrac{1}{R(R-1)^2} \,.
\end{equation}
Here we should assume that $R$ is sufficiently large since the truncated
expansion is a good approximation for $m$ only when $b$ is small.
In fact, $R > 10$ is reasonable, since $d_c(10)=0.9$ and $b_c(10)$ can be
numerically evaluated to give $b_c(10)\cong 0.00486$.
Using the approximation \eqref{eq:BAPPROX} one obtains $b_c(10)\approx 0.00123$.

From now on we shall split the analysis of the phenotypic model
according to which it is sub-critical, super-critical or critical.

\subsection{The Sub-critical Regime: Lethal Mutagenesis}

\enlargethispage{6mm}

The first consequence of our results is a proof, in the context of the phenotypic model
(provided one assumes that all effects are of purely mutational nature), of the
conjecture of \emph{lethal mutagenesis} of~\cite{BSW07}.
Recall that~\cite{BSW07} assumes that all mutations are either neutral or deleterious
and write the mutation rate $U=U_d+U_c$ where the component $U_c$ comprises the
purely neutral mutations and the component $U_d$ comprises the mutations with a 
deleterious fitness effect.
Let $R_{\mathrm{max}}$ denote the maximum reproductive capacity among all particles in
the viral population. 
The \emph{extinction criterion} proposed by~\cite{BSW07} states that a sufficient
condition for extinction is 
\begin{equation} \label{eq:LETHALMUT}
 \ee^{-U_d}R_{\mathrm{max}} < 1 \,.
\end{equation}
According to~\cite{BSW07,BSW08}, $\ee^{-U_d}$ is both the mean fitness level and also the
fraction of offspring with no non-neutral mutations.
In the absence of beneficial mutations the only type of non-neutral mutations are the
deleterious mutations and hence $\ee^{-U_d}=c=1-d$.
Since the evolution of the mean values depends only on the expected values of the
replication distribution $t_r$, it follows that $R_{\mathrm{max}}=R$.
That is, the extinction criterion \eqref{eq:LETHALMUT} is equivalent, in the context of 
the phenotypic model, to
\begin{equation} \label{eq:LETHALMUTALT}
 (1-d)\,R < 1 \,,
\end{equation}
which is exactly the condition for the model to be sub-critical.
In~\cite{BSW08} the authors suggest a modification of the extinction threshold 
\eqref{eq:LETHALMUT} that accounts for beneficial effects as long as they do not
couple the deleterious ones.
Assuming that the population size is large enough and that a large number of
individuals experience the full set of beneficial mutations they propose the following
threshold
\begin{equation} \label{eq:LETHALMUTEXT}
 \ee^{-U_d} \, R_{\mathrm{max}} \, (1+b) < 1 \,.
\end{equation}
Our formula \eqref{eq:PERTEXPAN} for the malthusian parameter allows us to propose a
generalization of the extinction threshold \eqref{eq:LETHALMUT} without any extra 
assumptions: if $b\neq 0$ is sufficiently small (up to order $\bigO(b^2)$)  
and
\begin{equation} \label{eq:LETHALMUTINTERACT}
 (1-d)\, R\, \left(1 + \dfrac{(R-1)\,d}{(1-d)^2} \, b \right) < 1
\end{equation}
then, with probability one, the population become extinct in finite time.

\ignore{
More precisely, $\bm{Z}=(Z^0,\ldots,Z^R)$ denotes the population at time $n$ then the
offspring at time $n+1$ is given by $\bm{M}\bm{Z}$ and thus
\[
\begin{split}
\ee^{-U_d} 
& = \dfrac{cZ^1+2cZ^2+\ldots+RcZ^R}{dZ^1+cZ^1+2dZ^1+2cZ^2+\ldots+RdZ^R+RcZ^R} \\
& = \dfrac{c}{d+c} \, \dfrac{Z^1+2Z^2+\ldots+RZ^R}{Z^1+2Z^2+\ldots+RZ^R} \\  
& = c = 1-d \,.
\end{split}
\]

\begin{theorem} \label{thm:CORLM}
Consider the simple phenotypic model with malthusian parameter $m=(1-d)R < 1$.
Then $\ee^{-U_d}=1-d$ and $R=R_{\mathrm{max}}$ and so the population become extinct
in finite time if the lethal mutagenesis criterion \eqref{eq:LETHALMUT} holds.
\end{theorem}
}
Observe that, when $b=0$ and $R$ fixed, the model becomes critical for $d=d_c(R)=1-1/R$.
In other words, for every fixed mean replication capacity $R$, if the deleterious
probability satisfies $d>d_c(R)$ then lethal mutagenesis occurs almost surely.
Thus one may consider a general \emph{critical probability} function $d_c(b,R)$ and 
in order to compute this probability as a function of $b$ and $R$ one must solve the
equation $m(b,d,R)=1$ for $d$ in terms of $(b,R)$.
This can be done by implicit differentiation of the equation $m(b,d,R)=1$:
\begin{equation} \label{eq:CRITPROBGEN}
\begin{split}
 d_c(b,R) & =(1-1/R)+b(R-1)^2+\bigO(b^2) \\
          & =d_c(R)+b(R-1)^2+\bigO(b^2) \,.
\end{split}
\end{equation}
Formula \eqref{eq:CRITPROBGEN} allows us to obtain a weak generalization of the
extinction criterion to the case where $b$ is non-zero and $d \approx d_c(R)$.
If $b\neq 0$ is sufficiently small and (up to order $\bigO(b^2)$)
\begin{equation} \label{eq:LETHALMUTGEN}
 (1-d)\,R < 1 - b\,R(R-1)^2
\end{equation}
then, with probability one, the population become extinct in finite time.
We summarize the discussion about extinction criteria in Table~\ref{tab:EXTCRIT}.

\renewcommand{\arraystretch}{1.5}

\begin{table}[!htb]
\begin{center}
\begin{tabular}{ccc} \hline
 Extinction Criterion & Conditions & Comments \\ \hline
 $(1-d)\, R < 1$ & & The \emph{lethal mutagenesis criterion (LMC)} \\
  (no beneficial mutations) & & of \cite{BSW07} with $(1-d)=\ee^{-U_d}$ \\[2mm]
 $(1-d)\,R\,(1+b)$ & $b \approx 0$, & Generalization of the LMC proposed in \\
  (uniform beneficial mutations) & $N\to\infty $ 
 & \cite{BSW08} ($N=$population size)  \\[2mm]
 $(1-d)\, R\, \left(1 + \dfrac{(R-1)\,d}{(1-d)^2} \, b \right) < 1$ 
 & $b \approx 0$ & Branching process generalization \\
 (strong version) & & obtained from perturbative expansion  \\[2mm]
 $(1-d)\, R < 1 - b\,R(R-1)^2$ 
 & $b \approx 0$, & Branching process generalization\\
 (weak version) & $d\approx d_c(R)$ & solving implict equation $m(b,d,R)=1$ \\
 \hline
\end{tabular}
\caption{\label{tab:EXTCRIT} Extinction criterion and its generaliations.
The role of beneficial mutations $b$ in several forms.} \vspace{-5mm}
\end{center}
\end{table}

\renewcommand{\arraystretch}{1}

The main conclusion here is that the existence of lethal mutagenesis depends on
``genetic components'' (mutational rates) and other additional deleterious effects
(host driven pressures intensifications), as well as on strict ``ecological
components'', namely, the maximum replication capacity of the particles in the
population and on the initial population size.
As a result the viral population may reach extinction by increasing the number of
deleterious mutations per replication cycle, by decreasing the value of $R$ in the
population or by a combination of the two mechanisms. 
The mutational strategy is the basis of treatments using mutagenic drugs 
(see~\cite{CCA01}) that induce errors in the generation process of new
viral particles reducing their replication capacity. 
A straightforward consequence of formulas \eqref{eq:LETHALMUTALT}, 
\eqref{eq:LETHALMUTINTERACT} and \eqref{eq:LETHALMUTGEN} is that a single particle
showing the maximum replication capacity $R$ is able to rescue a viral population 
driven to extinction by mutagenic drugs.
If it is assumed that RNA virus populations correspond to a swarm of variants with
distinct replication capacities, for a therapy to become effective it is important 
that it will eliminate the classes represented by particles with 
\emph{highest replication capacities}.    
In other words, the higher the replication capacity of the first particles infecting
the organism the larger should be the number of deleterious mutations (or effects), 
that is, the larger should be the drug concentration.
This can be a clear limitation for treatments based on mutagenic drugs.

\subsection{The Super-critical Case: Relaxation and Equilibrium}

In the super-critical regime, the population grows at a geometric pace indefinitely.
Nevertheless, there are two distinct phases that occur during this growth:
a transient phase (``relaxation''or ``recovery time'') and a dynamical stationary phase.

\subsubsection{Relaxation towards equilibrium}

An important question concerning the adaptation process of a viral population to the
host environment is the typical time needed to achieve the equilibrium state. As the
equilibrium is characterized by constant mean replication capacity an obvious criteria
to measure the time to achieve equilibrium would be by the vanishing of the variation 
of this variable as used in other studies \citep{ALM09}. 
Nevertheless, this method is clearly subjected to the limitations of numerical accuracy
with evident drawbacks if one wants a sharp and universal criterion to differentiate
populations from the point of view of how fast a population can be typically stabilized
in a organism. 

Viral populations are commonly submitted to transient regimes. 
As pointed out earlier the infection transmission process represents the passage of
a small number of particles from one organism to another in such a way that in this
process the viral population is submitted to a subsequent \emph{bottle-neck effect}
during spreading of viruses in the host population. 
In order to approach the problem of relaxation after a bottleneck process in a more
sound basis the natural quantity to be considered is the characteristic time derived
from the decay of the mean auto-correlation function. 
When the temporal mean correlation function $C(n)$ is of the form $\exp(-n \alpha)$ 
the \emph{decay rate} is defined as the parameter $\alpha$.   
The natural way to define the \emph{characteristic time} $T$ to achieve equilibrium
is by setting $T=1/\alpha$.

The \emph{temporal mean correlation function} $C(n)$ may be calculated by considering
recursive applications of the mean matrix $\bm{M}$ on the initial population:
$\bm{Z}_0^{\mathrm{t}}\,\bm{M}^n \bm{Z}_0$.
Since we are looking for bounds on the mean correlation function, it is enough
to consider the canonical initial populations $\bm{Z}_{0}=\bm{e}_r$ and use 
the Perron-Frobenius theorem (see \cite[p. 38]{H63} or \cite[p. 185]{AN72}) to write
\[
 \bm{Z}_0^{\mathrm{t}}\,\bm{M}^n \bm{Z}_0=m^n\big(u_R(r)v_R(r)\big) 
 + \bigO(\varrho_0^n) \,,
\]
where $\bm{u}_R$ is the \emph{normalized right eigenvector} and $\bm{v}_R$ is
\emph{normalized left eigenvector} (see the appendix for more details) corresponding 
to $m$ and $0<\varrho_0<m$.
Moreover, since we are assuming that $m>1$, it follows that 
$0<\tfrac{\bigO(\varrho_0^n)}{m^n}<1$ and $\tfrac{\bigO(\varrho_0^n)}{m^n}\to 0$
as $n$ goes to infinity.

Define the \emph{type} $r$ of an initial population $\bm{Z}_0$ as the largest
replicative class $r$ represented in that population.
In general, the  mean correlation function will depend only on the type $r$ of 
the initial population and will be denoted by $C_r(n)$.
Therefore, the mean correlation function is typically exponential and given by 
\[
 C_r(n)=K_r(n)\exp\big(-n\log(m)\big) \,,
\]
where $K_r(n)$ ($r=0,\ldots,R)$ is given by
\[
 K_r(n)=\dfrac{1}{\bm{u}_R(r)\bm{v}_R(r)+\tfrac{\bigO(\varrho_0^n)}{m^n}} \,.
\]
It is difficult, in general, to calculate an explicit expression of 
$\bm{u}_R(r)\bm{v}_R(r)$, but it is easy to see that 
\[
 \dfrac{1}{1+\bm{u}_R(r)\bm{v}_R(r)}< K_r(n) <\dfrac{1}{\bm{u}_R(r)\bm{v}_R(r)}
\]
and so the time dependence of $K_r$ is negligible.
Indeed, for $n$ sufficiently large the dominating factor of $C_r(n)$ is the exponential.

The lower bound for the mean correlation is attained when the type of the initial
population is $R$, that is,
\[
 K_R \approx 1+\bigO(b) \approx 1 \,.
\]
In this case the mean correlation function is
\[
 C_R(n) \approx \exp\big(-n\log(m)\big) \,,
\]
very similar to the case with no beneficial effects.
In fact, in this particular case $b=0$ and one has $K_r=1$, for every $r$, so it is
enough to consider the case $\bm{Z}_{0}=\bm{e}_R$, that is,
\[
 C(n)=\exp(-n\log(R(1-d))) \,.
\]
The difference between these two cases lies on the value of the largest eigenvalue
(the correlation decay rate) $m$.
As $m(b) > m(0)$ for the same values of $d$ and $R$ the correlation function decays
faster if beneficial effects are present.

In general, we have that (see the appendix)
\[
 K_{r} \approx \bigO(b^{-(R-r)})
 \qquad\text{for}\quad
 r=1,\ldots,R \,. 
\]
Hence the mean correlation function may be written as
\[
 C_{r}(n) \approx \dfrac{\kappa_r}{b^{R-r}}\,\exp\big(-n\log(m)\big) \,,
\]
where $\kappa_r=\kappa_r(d,R)$.

The conclusion here is that when the type of the initial population is $r$ there
is a ``delay effect'' on the decay of the mean correlation of order $b^{-(R-r)}$
relative to the fastest decay (lower bound) $C_R(n)$.
The slowest possible decay (upper bound) is attained when the type of the
population is $r=1$, which have a delay of order $b^{-(R-1)}$.
This ``delay effect'' on the mean correlation appears in simulations as
``jumps'' of magnitude $(1-d)$ of the mean replicative capacity, somewhat
reminiscent of the ``punctuated equilibrium scenario''.

Observe that the closer is the parameter $d$ to its critical value $d_c$, the longer 
is the time needed to achieve equilibrium.
The clearest way to characterize the time behavior of the viral population at or around
the critical point is through the typical time $T$ to approach equilibrium derived from
the decay of correlations described above.
The expression $T=1/\log(m)$ shows that at the critical point the equilibrium
state is never reached, i.e., the decay to equilibrium is at least non-exponential. 
A \emph{scaling exponent} characterizing the behavior of $T$ in the neighborhood of the
critical point $d_{c}$ can be easily obtained.
The expansion around $d_{c}$ gives 
\[
 T \approx |m-1|^{-1}=\dfrac{1}{m'(d_c)} \,|d-d_c|^{-1} \,.
\]
In the particular case where $b=0$ we have
\[
 T \approx (1-d_{c})\,|d-d_c|^{-1} \,,
\]
and hence $1/m'(d_c)=(1-d_c)+\bigO(b)$.

\ignore{
In fact, we have the expansion
\[
 m(d)-1=m'(d_c)(d-d_c)+\bigO\big((d-d_c)^2\big) \,.
\]
The we find
\[
 m'(d)=-R(1-(R-1)b/(1-d)+\bigO(b^2) \,.
\]
and
\[
 1/m'(d_c)=-(1/R+R(R-1)b+\bigO(b^2)
          =-(1-d_c+b(R-1))+\bigO(b^2) \,.
\]
Also we have the expansion around $d_c(b_c)1-b_c$:
\[
 m'(d)=-R+R^2(R-1)^3+R^3(R-1)^5 \,|d-d_c|^{-1}
\]
}

One possible application of this result relates to the very initial phase of the
infection process.
If one considers that during this phase the host immune system has not been yet
stimulated against the virus, one can assume that the deleterious effects would
be solely represented by the viral intrinsic mutation rates.
Therefore, the largest the value of $R$, i.e., the largest the replication capacity
of the initial viral particle, the fastest the decay of the progeny auto-correlation.
Intuitively the parameter $R$ defines the degree of virulence of the infection during 
the early stage of the infective process.

\subsubsection{The Dynamical Stationary State}

According to the ``Malthusian Law of Growth'' it is expected that a super-critical
branching process will grow indefinitely at a geometric rate proportional to $m^n$,
that is, $\bm{Z}_n \approx m^n \,\bm{W}_n$, where $\bm{W}_n$ is a random vector.
That is, the normalized random vector $\bm{W}_n=\bm{Z}_n/m^n$ posses a finite limit
when $n \to \infty$.
More precisely, there exists a random variable $\bm{W} \neq 0$ such that,
with probability $1$,
\[
 \lim_{n\to\infty} \bm{W}_n=\bm{W} \,.
\]
Moreover, $\bm{W}=W \bm{u}$ where $\bm{u}$ is the normalized right eigenvector
corresponding to the malthusian parameter $m$ and
\[
 \Expec(W|\bm{Z}_0)=\langle\bm{v}, \bm{Z}_0\rangle \,,
\]
where $\bm{v}$ is the left eigenvector corresponding to the malthusian parameter $m$.
The meaning of this result is that the total size of the population divided by $m^n$,
converges to a random vector, but the relative frequencies of the classes of 
particles approach fixed limits.
In fact, \citep{KLPP94} formalized this statement as a limit theorem
(no averages here!): 
\begin{equation} \label{eq:KURTZ}
 \lim_{n\to\infty} \,\dfrac{\bm{Z}_n}{|\bm{Z}_n|}=\bm{u} 
 \qquad\text{(almost surely).}
\end{equation}
This result is useful in computational simulations of the model, since one may find an
approximation to the ``deterministic part'' of $\bm{W}$ by sampling the population and
computing the relative frequencies of replicative classes.

\ignore{
In general, combining \eqref{eq:KURTZ} with the Perron-Frobenius theorem gives
\begin{equation} \label{eq:KPF}
 \lim_{n\to\infty} \,\dfrac{\bm{Z}_n}{|\bm{Z}_n|}=
 \lim_{n\to\infty} \,\dfrac{\langle\bm{Z}_n\rangle}{|\langle\bm{Z}_n\rangle|}=
 \bm{u} \,.
\end{equation}
Note that the approach adopted in \cite{CACM11,CCMA11,C11} relates only to the
second equality involving the limit of mean values of equation \eqref{eq:KPF}.
By explicitly considering the microscopic model as a multivariate branching 
process the equality of the two limits in equation \eqref{eq:KPF} is guaranteed.
}

Recall that the normalized right eigenvector $\bm{u}_R=(u_R(0),\ldots,u_R(R))$
corresponding to the malthusian parameter $m$ is positive and is normalized so that
$\sum_r u_R(r)=1$, therefore, it may be seen as a probability distribution on the set
of classes $\{0,\ldots,R\}$.
It is called the \emph{mean fitness distribution} of the replicative classes of
the viral population.

Using the same perturbative techniques as before, it is possible to compute the
perturbation expansion of the right eigenvector corresponding to the malthusian 
parameter.
We are interested in the right eigenvector $\bm{u}_R$ associated to the
malthusian parameter $m$.
The first order expansion of the normalized right eigenvector $\bm{u}_R$ 
associated to $m$ may be written as
\begin{equation} \label{eq:EIGMALTHNORM}
 \bm{u}_R=\bm{u}_R^0 + b \,\sum_{k=0}^{R} \beta_k \bm{u}_k^0 + \bigO(b^2) \,,
\end{equation}
where the terms $\beta_k$ are functions of $d$ and $R$.
The explicit expressions of the terms $\beta_k$ are somewhat cumbersome, 
however we still can use the abstract formula to gain some insight about $\bm{u}_R$.

\ignore{
When the phenotypic model, with $b=0$ is super-critical and is initialized with 
exactly one particle in the class $r$ ($Z_0^r=1$) it is equivalent to a
process with $R=r$ replicative classes since in the absence of beneficial mutations
it is not possible to increase the replicative capability of a particle.
Note that when $R=0,1$ the process is always sub-critical.
When $R\geqslant 2$ the malthusian parameter is $m=\lambda_R=Rc=R(1-d)$ can be greater
than $1$ and has a corresponding normalized right eigenvector $\bm{u}=(u_0,\ldots,u_R)$,

\begin{theorem} \label{thm:STATIONARYSTATE}
If the simple phenotypic model is super-critical with malthusian parameter $m=R(1-d)$ 
and starts with at least one particle of class $R$ then, in the long run, the relative
number of particles in each class reaches a stable stationary dynamical state and is
(up to a random scalar perturbation) distributed according to the 
\emph{Binomial Distribution:} $\mathrm{binom}(k;R,1-d)$, where $k=0,\ldots,R$ are the
replication classes.
\end{theorem}
\begin{proof}
This is consequence of Kesten-Stigum results about the asymptotic behaviour of 
decomposable super-critical multitype branching processes (see \cite{KS66a,KS66b} for
the case of indecomposable multitype branching processes and \cite{KS67} for the case
of a general decomposable multitype branching processes) and the computation of the
normalized right eigenvector associated to the malthusian parameter $m=R(1-d)$
given by equation~\eqref{eq:NREIGEN}. \qed
\end{proof}
}

The normalized right eigenvector $\bm{u}_R^0$ of $\bm{M}_0$ is
\[
 \bm{u}_R^0(r)=\mathrm{binom}(r;R,1-d)=\tbinom{R}{r} \, (1-d)^r \, d^{R-r} \,.
\]
From these formulas one immediately obtains, for $b=0$:
\begin{itemize}
\item The \emph{mean replication capacity} $\Expec(\bm{u}_R^0)=R(1-d)$.
\item The \emph{phenotypic diversity} $\Varia(\bm{u}_R^0)=Rd(1-d)$.
\end{itemize}
In general, the \emph{mean replication capacity} of a viral population is given by
\begin{equation} \label{eq:MRC}
 \Expec(\bm{u}_R)=\sum_{k=0}^R k \, u_R(k)=m \,.
\end{equation}
Indeed, the malthusian parameter may be calculated as
\[
 m=\lim_{n\to\infty} \dfrac{|\langle\bm{Z}_{n+1}\rangle|}{|\langle\bm{Z}_n\rangle|}
  =\lim_{n\to\infty} \dfrac{|\bm{M}\langle\bm{Z}_n\rangle|}{|\langle\bm{Z}_n\rangle|} \,.
\]
Now it is easy to see that $|\bm{M}\langle\bm{Z}_n\rangle|=\sum_k k Z_n^k$ and hence
by the Perron-Frobenius theorem, equation \eqref{eq:MRC} follows immediately.
In particular, when $b\neq 0$ is small enough:
\[
  \Expec(\bm{u}_R)=R(1-d)+\dfrac{(R-1)\,d}{1-d}\,R\,b+\bigO(b^2) \,.
\]

\ignore{
The calculation is as follows:
\[
 m=\lim_{n\to\infty} \dfrac{\sum_k k \, Z_n^k}{|\bm{Z}_n|}
  =\sum_{k=0}^R k \lim_{n\to\infty} \dfrac{Z_n^k}{|\bm{Z}_n|}
  =\sum_{k=0}^R k \,u_R(k) \,.
\]
}

Using \eqref{eq:EIGMALTHNORM} one may write the \emph{phenotypic diversity} of a
viral population as:
\begin{equation} \label{eq:VAREXP}
 \Varia(\bm{u}_R) = Rd(1-d) + \left(2 \sum_{k=0}^{R} \beta_k\,
 \mathbf{Cov}(\bm{u}_R^0,\bm{u}_k^0)\right)\, b + \bigO(b^2) \,.
\end{equation}
It is possible to estimate the magnitude of the first order perturbation term 
in eq.~\eqref{eq:VAREXP}:
\[
 \Varia(\bm{u}_R) \approx 
 Rd(1-d) \bigg(1+8 b \dfrac{R-1}{R}\dfrac{dR-1+(1-d)^R}{d^2}\bigg).
\]

It is well accepted that the phenotypic diversity is an important characteristic
of the viral population intuitively related to the idea of population
robustness \citep{FAW06,EWOL07}.
In fact, a homogeneous population would be less flexible from the point of view 
of adaptation. 
The variance associated with the stationary state can be seen as a measure of
phenotypic diversity.
It shows that, in the case where $b=0$, the maximum value of the phenotypic diversity
$R/4$ is reached if $d=1/2$ for any value of $R$.
For $b \neq 0$ it can be shown numerically, that there is always a value $d_{\mathrm{max}}$
of $d$ for which the population attains maximal phenotypic diversity.
This value $d_{\mathrm{max}}$ is a decreasing function of $b$.
For $b \approx 0$ this value is $d_{\mathrm{max}} \lesssim 1/2$.

It is interesting to note that experimental measurements of $c=1-d$ are close to $0.5$
\citep{SME04,CCS09}, suggesting that viral populations follow a principle of maximal 
phenotypic diversity.

As far as the experimental situation of the phenotypic diversity is concerned, one of
the first attempts to  experimentally measure the phenotypic diversity was in
\cite{D45}, where the total ``burst size'' of a progeny from phage-infected bacterial
cells was estimated.
More recently, measurements of the phenotypic diversity \emph{in vitro} generated
by single viral particles infecting individual cells \citep{ZYY09,TY12} revealed 
a broad distribution of virus yields ($50$ to $8000$ progeny virus particles).
One may regard these results as indication that the replicative capacity of a 
virus from a particular replicative class is characterized by a 
\emph{fitness distribution} obtained as the distribution of progeny produced 
by representative particles from that same replicative class.
Although these authors did not went further and investigated if different particles 
form a viral population have different fitness distributions their results suggest 
that a fitness distribution over a viral population may resemble the 
\emph{mean fitness distribution} of the replicative classes obtained from 
the phenotypic model. 
In this direction, further carefully designed experiments aiming at determine the 
mean fitness distribution of a viral population would be welcome.

\subsection{Criticality and Eigen's Error Threshold}

The relation between multitype branching processes and Eigen's theory of evolution of
macromolecules~\citep{E71} has been established by~\cite{DSS85}.
In this work it is shown that one may canonically associate a certain difference
equation to a discrete time multitype branching process through its mean matrix.
This deterministic dynamical system is called \emph{Eigen's selection equation},
due to its similarity to the phenomenological equation describing the kinetics of
self-reproducing molecules in a dialysis reactor.
Moreover, the selection equation is essentially equivalent to a linear difference
equation $\bm{x}_{n}=\bm{M}^n\bm{x}_{0}$, where $\bm{M}$ is the mean matrix of the
branching process.
Note that this equation is formally identical to the equation for the evolution of the
mean values of branching process (see eq.~\eqref{eq:MEANEVOL1}).
Hence the selection equation may be seen as a mean field limit equation associated
branching process.
In particular, when the process does not become extinct, part of its asymptotic 
behaviour, namely, the relative frequencies the classes, can be recovered by the
selection equation. 
Moreover, it is shown that a super-critical branching process displays ``freezing in'' 
of initial fluctuations, that is, the \emph{coefficient of variation} of the process
vanishes asymptotically almost sure. 
In this sense, if the population is infinite and one waits long enough before starting
observation, the deterministic model is fairly reliable because fluctuations are small. 
Nonetheless, when considering finite populations, i.e., finite size samples of each
generation, the deterministic approximation is only part of the story, since the
fluctuations contain the ``out-of-equilibrium'' characteristics of the system.

The use of branching processes links the concept of \emph{error threshold} with
that of criticality.
If we switch to the genotypic view of \cite{DSS85} and suppose that there are no
back-mutations ($b=0$) then we can consider polymers with chain length of $\nu$
nucleotides and that there is a fixed probability $p$ for copying a single nucleotide 
correctly, that is, $c=p^\nu$.
The formula for the malthusian parameter ($m=cR=p^\nu R$) together with the 
criticality condition ($m=1$) gives the \emph{stochastic error threshold} 
of~\citep[p. 254, eq. (47)]{DSS85}:
\begin{equation} \label{eq:ERROR1}
 \nu_{\mathrm{s}}~=~\dfrac{\;\ln R}{-\ln p} \,. 
\end{equation}
Since the lethal mutagenesis criterion, in the context of branching processes,
is exactly the criticality condition, the stochastic error threshold refers 
to the probability of extinction.
This can be compared with the \emph{deterministic error threshold}
(see \cite{E71,ES79,SS82}):
\begin{equation} \label{eq:ERROR2}
 \nu_{\mathrm{d}}~=~\dfrac{\;\ln \sigma}{-\ln p} \,, 
\end{equation}
where the parameter $\sigma$ is the \emph{superiority} of the master sequence
(see \cite{DSS85}).
The deterministic error threshold is based on the condition that the error-free
productivity of the master molecular species becomes equal to the mean total productivity 
of all other species.
The condition to replicate with a fidelity above the error threshold will always be valid
for the master species provided the mutation terms of all other species are sufficiently
small.
Beyond that point, the master species is no longer maintained: the population has reached
the \emph{error catastrophe}~\citep{B05}.
Moreover, the demand to operate above the stochastic threshold is always
a stronger condition than the corresponding requirement of the deterministic equation.

\pagebreak

It is interesting to note that at the critical point the dynamics of the branching process
relax to equilibrium according to the typical algebraic decay in time.
This property raises the question (unapproachable in the present context) if the classes
in the critical population are so strongly correlated in such a way that the whole
population behaves as if it is composed be only one replicative class.
This behaviour reminds one of the basic properties of the \emph{error threshold} 
in Eigen's theory of macromolecular evolution~\citep{E71}.

Preliminary results concerning the dynamics of fluctuations (\cite{ABJ13,D11}) indeed 
show that the time variation of the numbers of particles in each separated class follows 
a pattern such that the variation observed in one class is rigorously the same observed
in all the others.
This indicate a high level of correlation between the classes in complete analogy 
with critical phenomena of many physical systems.
It is worth to note that, in fact, there is a correspondence between Eigen's model 
and the equilibrium statistical mechanics of a certain inhomogeneous Ising system 
(see~\cite{L87}).

\section{Outlook}
\label{sec:CO}

Modeling the dynamics of viral populations by means of multivariate branching 
processes does not take into account many molecular/microscopic details of the 
replicative process, nevertheless, it is remarkable that the provided description 
at the population/macroscopic scale is very appealing and well fitted to various
aspects of the observed phenomenology of viral systems.
This leads to the conclusion that we are probably far from exhaust the analytical
and predictive power of branching processes as a mathematical tool for the study of
viral populations.
In fact, many aspects of the problem like (i) the role of finite size effects,
(ii) the role of different infected cell types for the problem of evolution of
viral populations in multi-cellular hosts, (iii) the relation between quiescent
infected cells with branching process with singular components and (iv) the
descriptive limits of discrete versus continuous time branching processes, are
still to be considered by future investigations.
Also it is worth to mention that most of the quantitative results and all of the
qualitative observations described here can be easily reproduced by computer simulation
of the model (see Castro~\cite{D11} and manuscript in preparation~\cite{CABJ11}). 

%{\color{Red}
Finally, it is important to emphasize that the branching process analysis carried 
out here revolves around the malthusian parameter, which is based on certain
critical assumptions regarding environmental constraints, for instance, that
the environment (host) does not change dramatically.
The significance of this macroscopic measure in studies of evolutionary processes has
been challenged in a series of articles (see~\cite{D05,KD05,ZD05}).
When the environment is strongly perturbed, another parameters become important,
as the ``evolutionary entropy'', which is a measure of the variability in the age at
which individuals produce offspring and die (see~\cite{KD05,ZD05}). These parameters
will play a role in studies of the extinction process. 

\vspace{\fill}

\begin{acknowledgements}
The authors would like to thank C.O.~Wilke for some pertinent comments on
previous version of this work.
FA wish to acknowledge the support of CNPq through the grant PQ-313224/2009-9.
FB recieved support from the Brazilian agency FAPESP. 
DC received financial support from the Brazilian agency CAPES. 
\end{acknowledgements}

\appendix

\section*{Appendix: Spectral Perturbation Theory}

In order to carry out the perturbative calculations we need to solve the spectral
problem for the matrix
\[
\bm{M}_0=\begin{pmatrix}
 0 & d     &  0     & \ldots & 0 \\
 0 & (1-d) & 2d     & \ldots & 0 \\
 0 & 0     & 2(1-d) & \ldots & 0 \\
 0 & 0     &  0     & \ldots & 0 \\[-2mm]
\vdots     & \vdots & \vdots & \ddots & Rd \\
 0 & 0 & 0 & \ldots & R(1-d)
\end{pmatrix} \,.
\]
The eigenvalues $\lambda_r^0$ of the mean matrix $\bm{M}_0$ are:
\[
 \lambda_r^0=rc=r(1-d) \qquad r=0,\ldots,R \,.
\]
In particular, the \emph{malthusian parameter} is the largest positive eigenvalue
\[
 m^0=\lambda_R^0=Rc=R(1-d) \,.
\]
The normalized left and right eigenvectors, $\bm{v}_r^0$ and $\bm{u}_r^0$, associated
to the eigenvalue $\lambda_r^0$, satisfy:
\[
 (\bm{v}_s^0)^{\mathrm{t}}\bm{u}_s^0=1 
 \quad\text{and}\quad
 \bm{1}^{\mathrm{t}}\bm{u}_s^0=1 \,.
\]
Writing $\bm{v}_r^0$ and $\bm{u}_r^0$ in components as 
\[
\begin{split}
 \bm{v}_r^0 & =\big(v_r^0(0),v_r^0(1),\ldots,v_r^0(R)\big) \,, \\
 \bm{u}_r^0 & =\big(u_r^0(0),u_r^0(1),\ldots,u_r^0(R)\big) \,,
\end{split}
\]
one has that
\[
 v_r^0(k)=\left\{\begin{array}{c@{\quad\text{for}\quad}c} 
 0 & k=0,\ldots,r-1 \,, \\[3mm]
 \dfrac{-d(r+1)}{(1-d)^{r+1}} & k=r+1\\[3mm]
 \dfrac{((-1) \,d)^{k-r}}{(1-d)^{k}} & k=r,r+2,\ldots,R \,.
 \end{array}\right.
\]
and
\[
 u_r^0(k)=\left\{\begin{array}{c@{\quad\text{for}\quad}c}
 \tbinom{r}{k} \, (1-d)^k \, d^{r-k} & k=0,\ldots,r \,, \\[2mm]
 0 & k=r+1,\ldots,R \,.
 \end{array}\right.
\]

Now we write the eigenvalue $\lambda_r$ of $\bm{M}$ as a function of the parameter $b$,
expanded as a power series
\[
 \lambda_r = \lambda_r^0 + \lambda_r^1 b + \lambda_r^2 b^2 + \ldots \,,
\]
where $\lambda_r^0$ is the corresponding eigenvalue of $\bm{M}_0$.
Clearly, $\lambda_r(b)\to\lambda_r^0$ as $b\to 0$.
The higher order perturbation coefficients $\lambda_r^i$, ($i=1,2,3,\ldots$) 
are written in terms of all the eigenvalues $\lambda_s^0$ ($s=0,\ldots,R$) 
of $\bm{M}_0$ and their associated left and right normalized eigenvectors 
$\bm{v}_s^0$ and $\bm{u}_s^0$ and the perturbation matrix
\[
\bm{P}=\begin{pmatrix}
 0 &  0 &  0 &  0 &  0 & \ldots &  0 & 0 \\
 0 & -1 &  0 &  0 &  0 & \ldots &  0 & 0 \\
 0 &  1 & -2 &  0 &  0 & \ldots &  0 & 0 \\
 0 &  0 &  2 & -3 &  0 & \ldots &  0 & 0 \\
 0 &  0 &  0 &  3 & -4 & \ldots &  0 & 0 \\[-2mm]
 \vdots &  \vdots &  \vdots & \vdots & \vdots & \ddots & \vdots & \vdots \\ 
 0 &  0 &  0 &  0 & 0  & \ldots & -(R-1) & 0 \\
 0 &  0 &  0 &  0 & 0  & \ldots &  (R-1) & 0
\end{pmatrix} \,.
\]
Note that the operator norm of $\bm{P}$ is $\|\bm{P}\|=2(R-1)$ and so the
magnitude of the perturbation is $2b(R-1)$.

The general expressions for the first and second order coefficients of the perturbation
expansion of an eigenvalue are
\[
\begin{split}
 \lambda_r^1 & =(\bm{v}_r^0)^{\transp}\bm{P}\bm{u}_r^0 \,, \\
 \lambda_r^2 & =\sum_{s\neq r} \dfrac{\big[(\bm{v}_s^0)^\transp\bm{P}\bm{u}_r^0\big] \,
 \big[(\bm{v}_r^0)^\transp\bm{P}\bm{u}_s^0\big]}{\lambda_r^0-\lambda_i^0} \,.
\end{split}
\]
We want to use these formulas to compute a perturbation approximation 
(for $b$ around $0$) of the malthusian parameter $m(b)$ of $\bm(M)$.
Recall that, $m(0)=m^0=R(1-d)$ and the corresponding left and right eigenvectors 
are given by
\[
\begin{split}
 \bm{v}_R^0 & =1/(1-d)^R(0,\ldots,0,1) \,, \\
 \bm{u}_R^0 & =(u_R(0),\ldots,u_R(R)) \,,
\end{split}
\]
with $u_R(k)=\mathrm{binom}(k;R,1-d)$.
The first order coefficient of the malthusian parameter is
\[
 m^1 = (\bm{v}_R^0)^{\transp}\bm{P}\bm{u}_R^0 = \dfrac{(R-1)u_R(R-1)}{(1-d)^R}
 = \dfrac{R(R-1)d}{1-d} \,.
\]
For the second order coefficient we also need 
\[
 \bm{v}_{R-1}^0=1/(1-d)^{R-1}(0,\ldots,0,1,-Rd/(1-d)) \,.
\]
The second order coefficient is
\[
\begin{split}
 m^2 & = \sum_{i=0}^{R-1} 
 \dfrac{\big[(\bm{v}_i^0)^\transp\bm{P}\bm{u}_R^0\big] \,
 \big[(\bm{v}_R^0)^\transp\bm{P}\bm{u}_i^0\big]}{\lambda_R^0-\lambda_i^0} \\[3mm]
 & = \dfrac{\big[(\bm{v}_{R-1}^0)^\transp\bm{P}\bm{u}_R^0\big]\,
     \big[(\bm{v}_R^0)^\transp\bm{P}\bm{u}_{R-1}^0\big]}{(1-d)} \\[3mm]
 & = -\dfrac{R(R-1)^2(1+\tfrac{R}{2}\,d)d}{(1-d)^3} \,.
\end{split}
\]

\ignore{
It is interesting to compare the perturbation expansion of $m$ with the exact formula
in the first non-trivial case $R=2$, where it is possible to find the eigenvalues
explicitly:
\[ 
m_{(R=2)}= 2(1-d)-\dfrac{1}{2}\big((1-d)+b-\sqrt{(1-d)^2+b^2+b(2+6d)}\,\big) \,.
\]
}

Analogous perturbative formulas exist for the perturbation expansion of the left 
and right eigenvector corresponding to an eigenvalue $\lambda_r$.
The left and right eigenvector $\bm{v}_r$ and $\bm{u}_r$ of the matrix $\bm{M}$
associated with the eigenvalue $\lambda_r$ can be written as a function of the parameter
$b$, expanded as a vector-valued power series: 
\[
\begin{split}
 \bm{v}_r & =\bm{v}_r^0
 + b\,\sum_{s \neq r} A_{r,s}^1\bm{v}_s^0
 + b^2\,\sum_{s \neq r} A_{r,s}^2\bm{v}_s^0
 +\ldots \,, \\
 \bm{u}_r & =\bm{u}_r^0
 + b\,\sum_{s \neq r} B_{r,s}^1\bm{u}_s^0
 + b^2\,\sum_{s \neq r} B_{r,s}^2\bm{u}_s^0
 +\ldots \,,
\end{split}
\]
where $\bm{v}_r^0$ and $\bm{u}_r^0$ are, respectively, the left and right eigenvectors
of the matrix $\bm{M}_0$ associated to the eigenvalue $\lambda_r^0$. 
The perturbation terms $A_{r,s}^i$ and $B_{r,s}^i$ with $i=1,2,3,\ldots\,$, can be
written in terms of the eigenvalues of $\bm{M}_0$ and their associated left and right
eigenvectors and the perturbation matrix $\bm{P}$.
Observe that when $b\to 0$ we get the normalized left and right eigenvectors
$\bm{v}_R^0$ and $\bm{u}_R^0$ associated to the dominant eigenvalue $m^0=(1-d)R$
of the mean matrix $\bm{M}_0$.

We are interested in the normalized eigenvectors $\bm{v}_R$ and 
$\bm{u}_R$ associated to the malthusian parameter $m=(1-d)R$
\[
 \bm{v}_R=\bm{v}_R^0 + b \,\sum_{k=0}^{R} \alpha_k \bm{v}_k^0 + \bigO(b^2) \,,
\]
with $\alpha_k=A_{R,k}^1$, ($0\leqslant k\leqslant R-1$) and 
$\alpha_R=-\sum_{k=0}^{R-1} \alpha_k$,
\[
 \bm{u}_R=\bm{u}_R^0 + b \,\sum_{k=0}^{R} \beta_k \bm{u}_k^0 + \bigO(b^2) \,,
\]
with $\beta_k=B_{R,k}^1$, ($0\leqslant k\leqslant R-1$) and 
$\beta_R=-\sum_{k=0}^{R-1} \beta_k$.

The first order coefficients are given by
\[
\begin{split}
 A_{R,s}^1 & = \dfrac{(\bm{v}_R^0)^{\transp}\bm{P}\bm{u}_s^0}{\lambda_R^0-\lambda_s^0} \,,\\
 B_{R,s}^1 & = \dfrac{(\bm{v}_s^0)^{\transp}\bm{P}\bm{u}_R^0}{\lambda_R^0-\lambda_s^0} \,,
\end{split}
\]
and the second order terms are given by
\[
\begin{split}
 A_{R,s}^2 & = \sum_{i=0}^{R-1}
 \dfrac{\big[(\bm{v}_R^0)^\transp\bm{P}\bm{u}_i^0\big] \,
 \big[(\bm{v}_i^0)^\transp\bm{P}\bm{u}_s^0\big]}
 {(\lambda_R^0-\lambda_s^0)\,(\lambda_R^0-\lambda_i^0)} \,, \\
 B_{R,s}^2 & = \sum_{i=0}^{R-1}
 \dfrac{\big[(\bm{v}_s^0)^\transp\bm{P}\bm{u}_i^0\big] \,
 \big[(\bm{v}_i^0)^\transp\bm{P}\bm{u}_R^0\big]}
 {(\lambda_R^0-\lambda_s^0)\,(\lambda_R^0-\lambda_i^0)} \,,
\end{split}
\]
for $s=0,\ldots,R-1$.

\ignore{
It is fairly easy to calculate the second order expansion of the (unormalized)
left eigenvector $\bm{v}_R$ associated to the malthusian parameter $m$:
\[
\begin{split} 
 \bm{v}_R = \bm{v}_R^0 
            & + b\,\dfrac{(R-1)}{(1-d)^2}\,\bm{v}_{R-1}^0 \\
            & - b^2\,\dfrac{(R-1)(1+(2R-3)\,d)}{(1-d)^3}\,\bm{v}_{R-1}^0 \\
            & + b^2\,\dfrac{(R-1)(R-2)}{2(1-d)^3}\,\bm{v}_{R-2}^0 
              + \bigO(b^3) \,.
\end{split}
\]
Using the previous computations it is easy to see that one has 
$A_{R,s}^1=0$ for $s=0,\ldots,R-2$ and
\[
 A_{R,R-1}^1
 = \dfrac{(\bm{v}_R^0)^{\transp}\bm{P}\bm{u}_{R-1}^0}{\lambda_R^0-\lambda_{R-1}^0}
 = \dfrac{(R-1)(1-d)^{R-1}}{(1-d)^{R+1}}=\dfrac{R-1}{(1-d)^2} \,.
\]
Likewise, for the second order terms one has $A_{R,s}^1=0$ for $s=0,\ldots,R-3$.
For $s=R-2$ one has
\[
\begin{split}
 A_{R,R-2}^2
 & = \dfrac{\big[(\bm{v}_R^0)^\transp\bm{P}\bm{u}_{R-1}^0\big] \,
   \big[(\bm{v}_{R-1}^0)^\transp\bm{P}\bm{u}_{R-2}^0\big]}
   {(\lambda_R-\lambda_{R-1})(\lambda_R^0-\lambda_{R-2}^0)} \\
 & = \dfrac{(R-1)(R-2)(1-d)^{R-2}}{2(1-d)^{R+1}} \\
 & = \dfrac{(R-1)(R-2)}{2(1-d)^3}
\end{split}
\]
and for $s=R-1$ one has
\[
\begin{split}
 A_{R,R-1}^2
 & = \dfrac{\big[(\bm{v}_R^0)^\transp\bm{P}\bm{u}_{R-1}^0\big] \,
   \big[(\bm{v}_{R-1}^0)^\transp\bm{P}\bm{u}_{R-1}^0\big]}
   {(\lambda_R-\lambda_{R-1})(\lambda_R^0-\lambda_{R-1}^0)} \\
 & = -\dfrac{(R-1)(1+(2R-3)\,d)}{(1-d)^3} \,.
\end{split}
\]
}

One may use the formula of $\bm{v}_R$ up to second order
\[
 \bm{v}_R = \bm{v}_R^0 + b\,\dfrac{(R-1)}{(1-d)^2}\,\bm{v}_{R-1}^0 + \bigO(b^2) 
\]
and the first order expansion $\bm{u}_R = \bm{u}_R^0 + \bigO(b)$ in order
to estimate the product of coefficients $u_R(r)v_R(r)$ (up to their leading order). 
For instance, it is easy to find that
\[
\begin{split}
 u_R(R)v_R(R) = 1+\bigO(b) & = \bigO(1) \,, \\
 u_R(R-1)v_R(R-1) & = \bigO(b) \,, \\
 u_R(R-2)v_R(R-2) & = \bigO(b^2) \,.
\end{split}
\]
In general, using a higher order expansion formula for $\bm{v}_R$, one concludes that
\[
 u_R(R-r)v_R(R-r) = \bigO(b^r) \,.
\]
This follows form the fact that the coefficient of order $b^r$ in the perturbation
expansion of $\bm{v}_R$ is a linear combination of the left eigenvectors
$\bm{v}_{R-1}^0,\ldots,\bm{v}_{r-R}^0$, all of which have zeros in the first $r-1$
components.

\providecommand{\noopsort}[1]{}\providecommand{\singleletter}[1]{#1}%

\end{document}